# Common Synaptic Input, Synergies, and Size Principle: Control of Spinal Motor Neurons for Movement Generation

François HUG[1,2], Simon AVRILLON[3], Jaime IBÁÑEZ[4], Dario FARINA[3]

[1] Université Côte d'Azur, LAMHESS, Nice, France
[2] Institut Universitaire de France (IUF), Paris, France
[3] Department of Bioengineering, Imperial College London, UK
[4] BSICoS, IIS Aragón, Universidad de Zaragoza, Zaragoza, Spain

Correspondence:
Prof. François Hug
Université Côte d'Azur, LAMHESS, Nice, France
E-mail: francois.hug@univ-nantes.fr

Prof. Dario Farina
Imperial College London, London, UK
E-mail: d.farina@imperial.ac.uk

**Acknowledgements**
François Hug was supported by a fellowship from the Institut Universitaire de France (IUF) and by the French government, through the UCAJEDI Investments in the Future project managed by the National Research Agency (ANR) with the reference number ANR-15-IDEX-01. Jaime Ibáñez was supported by a fellowship from "la Caixa" Foundation (ID 100010434) and from the European Union's Horizon 2020 Research and Innovation Programme under the Marie Skłodowska-Curie Grant Agreement No 847648 (fellowship code LCF/BQ/PI21/11830018). Dario Farina received support from the European Research Council Synergy Grant NaturalBionicS (contract #810346), the EPSRC Transformative Healthcare, NISNEM Technology (EP/T020970), and the BBSRC, "Neural Commands for Fast Movements in the Primate Motor System" (NU-003743).
The authors thank Andrew J. Fuglevand (University of Arizona) for his insightful comments on the manuscript.

# Abstract

Understanding how movement is controlled by the central nervous system remains a major challenge, with ongoing debate about basic features underlying this control. In this review, we introduce a new conceptual framework for the distribution of common input to spinal motor neurons. Specifically, this framework is based on the following assumptions: 1) motor neurons are grouped into functional groups (clusters) based on the common inputs they receive; 2) clusters may significantly differ from the classical definition of motor neuron pools, such that they may span across muscles and/or involve only a portion of a muscle; 3) clusters represent functional modules used by the central nervous system to reduce the dimensionality of the control; and 4) selective volitional control of single motor neurons within a cluster receiving common inputs cannot be achieved. We discuss this framework and its underlying theoretical and experimental evidence.

# 1. Introduction

Understanding how movement is controlled by the central nervous system (CNS) remains a major challenge, with ongoing debates about basic characteristics of its neural determinants. For example, the orderly recruitment of spinal motor neurons by size, as originally observed by Henneman in the 50s (Henneman, 1957), is still challenged (Menelaou & McLean, 2012; Kishore *et al.*, 2014; Azevedo *et al.*, 2020; Formento *et al.*, 2021; Marshall *et al.*, 2022). Moreover, the possible strategies used by the CNS to reduce the computational burden of coordinating several thousand motor neurons across many muscles during natural movements have been proposed and debated for decades (Bernstein, 1947; Tresch & Bizzi, 1999; d'Avella & Bizzi, 2005; Bizzi & Cheung, 2013; Latash, 2021). We propose that a fundamental feature of movement control is the distribution of correlated synaptic inputs across functional, rather than anatomical, groups of spinal motor neurons. Understanding the structure of these correlated inputs may provide an important framework for a better understanding of how the CNS achieves two apparent opposing goals: reducing the dimensionality of control and flexibly recruiting motor neurons to comply with various task constraints. Herein, we discuss this framework and the underlying theoretical and experimental evidence.

# 2. Inputs to motor neurons

*2.1. Common input control*

Each motor neuron transduces the synaptic input it receives into a series of action potentials that reach and excite an innervated group of muscle fibres, that is, the muscle unit. A key role of the CNS in movement generation is to provide appropriate inputs to populations of motor neurons. As the final common pathway (Sherrington, 1906), the motor neuron receives inputs from descending, spinal interneuronal, and sensory systems through thousands of synaptic connections. The net excitatory input that results from this synaptic bombardment can be considered an equivalent input current. A portion of this input is correlated across the motor neurons. We refer to this part of the input as common synaptic input and to the remaining portion as independent input.

Although both common and independent inputs to motor neurons theoretically contribute to the amount of force produced by the muscle, they have different effects on force modulation. Indeed, fine-tuning of the force level (i.e. modulation around a mean force) necessarily requires a concomitant and coherent activation of the muscle units (Negro *et al.*, 2009; Farina *et al.*, 2014). Therefore, volitional force modulation is mainly determined by common input, which results in common fluctuations in the discharges of motor neurons. Importantly, this does not necessarily imply that only common inputs are transmitted to motor neurons for force modulation, but instead that a minimum amount of common input should be present in natural behaviour to voluntarily modulate muscle force. Moreover, while the frequency range of the neural drive that influences muscle force is within approximately 10–12 Hz, common input at higher frequencies could also be non-linearly transmitted to lower frequencies (Watanabe & Kohn, 2015); therefore, the frequency content of the common input does not exactly replicate the bandwidth of the neural drive to the muscle. Nonetheless, even if non-linear transformations of common input in the neural drive may occur, the input still needs to be common to a sufficient number of motor neurons to influence the time-varying modulation of muscle force. Henneman's *size principle* imposes rigid control (Marshall et al., 2022) on motor neurons receiving common inputs. The size principle asserts that the physical dimensions of the soma and dendrites determine how readily a motor neuron is brought to spiking threshold. Smaller-sized motor neurons, which innervate weaker muscle units, possess a higher input resistance. Because of Ohm's law, changes in membrane potential will be greater in smaller motor neurons than in larger ones in response to equivalent synaptic currents. Consequently, across motor neurons receiving similar synaptic input, i.e. common input, the recruitment order of motor units should progress 'automatically' from those innervated by small neurons producing weak forces toward those innervated by large neurons exerting larger forces (Henneman, 1957). The combination of the size principle and common input is thought to be an effective way to reduce the computational load associated with controlling a large number of motor neurons (Henneman & Mendel, 1981). Importantly, such rigid control would be observed among the motor neurons receiving similar (i.e. common) input, regardless of the motor task. Conversely, motor neurons which receive different or independent inputs can, theoretically, be selectively recruited.

*2.3. Conflicting evidence for common input control*

Measurement of common inputs to motor neurons during natural motor tasks is not straightforward. Indeed, it is not possible to measure the synaptic currents to motor neurons; therefore, only indirect measures based on their output are possible. The correlation of motor unit discharge times (referred to as motor unit *[short-term] synchronisation*) (Sears & Stagg, 1976; Heckman & Enoka, 2012) has been widely used to infer the presence of common synaptic input to motor neurons. Nonetheless, the level of synchronisation between trains of action potentials of two motor neurons in their full bandwidth is not linearly proportional to the degree of common synaptic input to these motor neurons (de la Rocha *et al.*, 2007). Consequently, the absence of synchronisation cannot be considered conclusive evidence for the absence of a common input (Farina & Negro, 2015). This is mainly explained by the fact that a motor neuron typically undersamples its synaptic input due to its relatively low discharge rate (< 40 pulses/s for most muscles). This leads to a non-linear relationship between the input and the output signal, which is more pronounced at higher input frequencies (Negro & Farina, 2012; Farina & Negro, 2015). This problem can be mitigated by assessing the correlation between low-frequency oscillations of motor neuron discharge rates (Negro and Farina, 2012), a concept originally termed *common drive* (De Luca *et al.*, 1982). The strength of the common drive can be estimated by applying a low-pass filter to the motor neuron discharge times before assessing their correlation (De Luca & Erim, 1994; Semmler *et al.*, 1997).

Even though assessing whether motor neurons share a common input is challenging and influenced by factors that cannot be fully identified or compensated for, a combination of experimental results provides evidence of shared common inputs to motor neurons. Indeed, several studies have observed synchronisation or common drive to motor neurons innervating the same muscle (Schmied *et al.*, 1994; Semmler & Nordstrom, 1995). In addition, synchronisation of motor neurons has been reported across synergist muscles, for example, between the vastus lateralis and medialis (Mellor & Hodges, 2005), extensor carpi radialis longus and extensor carpi ulnaris (De Luca & Erim, 2002), and medial gastrocnemius and soleus (Gibbs *et al.*, 1995). The presence of such common input likely explains why, in a study where participants were provided with real-time feedback of the activity of pairs of motor neurons, they failed to volitionally control individual motor neurons (Bräcklein *et al.*, 2021).

Despite the aforementioned evidence for common input control, it is important to note that other studies have reported flexible control of motor neurons. Selective voluntary activation of single motor units has been suggested (Basmajian, 1963; Formento *et al.*, 2021) and notable exceptions to the size principle have been reported (Basmajian, 1963; Smith *et al.*, 1980; Desmedt & Godaux, 1981; Kishore *et al.*, 2014; Azevedo *et al.*, 2020; Marshall *et al.*, 2022). For example, the recruitment order of motor neurons innervating the human interosseous muscle can change based on movement direction (Desmedt & Godaux, 1981). Violation of the size principle was also observed during rapid paw shaking behaviour in cats (Smith et al., 1980) or during rapid escape behaviours in zebrafish, where the most excitable motor neurons were not recruited (Menelaou & McLean, 2012). In addition, Marshall et al. (2022) observed behaviour-dependent patterns of motor neuron recruitment during isometric tasks performed under different mechanical constraints in macaques. These observations are often cited as evidence for inverted, rather than size-based, recruitment of motor neurons. Therefore, they are often used to support the capacity of the CNS to selectively target individual motor neurons.

In conclusion, there is conflicting evidence regarding purely common input control. As discussed below, we contend that changing the scale at which we observe and interpret common input to spinal motor neurons may reconcile previous divergent interpretations.

## 3. A new conceptual framework for the distribution of common input to spinal motor neurons

The concept of common input is often discussed at the level of the motor neuron pool, which is defined as the ensemble of motor neurons that innervate a muscle. Therefore, it is assumed (either explicitly or implicitly) that the full pool of motor neurons innervating a muscle receives similar (common) inputs (Figs. 1A and B) (De Luca & Erim, 1994). This assumption is also made implicitly when studying muscle synergies (Fig. 2A), defined as functional units that generate a motor output by imposing a specific activation pattern on a group of muscles (d'Avella & Bizzi, 2005; Cheung & Seki, 2021). Indeed, these investigations on muscle synergies considered the muscle as the smallest functional unit of analysis (Giszter, 2015). Here, we introduce an alternative concept (Fig. 3A), where common inputs do not necessarily project to all motor neurons in a pool but to groups of motor neurons that pertain to different pools. Specifically, this framework is based on the following assumptions: 1) motor neurons are grouped into functional groups (clusters) based on the common inputs they receive; 2) clusters may significantly differ from the classical definition of motor neuron pools, such that clusters of motor neurons may span across muscles and/or involve only a portion of a muscle; 3) clusters represent functional modules used by the CNS to reduce the dimensionality of the control; and 4) selective volitional control of single motor neurons within a cluster receiving common inputs cannot be achieved. The composition and number of clusters may change flexibly to accommodate a variety of tasks and learn new motor skills. Importantly, in addition to this organisation, motor neurons may receive distinct proprioceptive feedback signals, and their intrinsic properties may be independently modulated through persistent inward currents (Heckman *et al.*, 2008). This framework assumes that the CNS does not control muscles, but rather controls functional clusters of motor neurons.

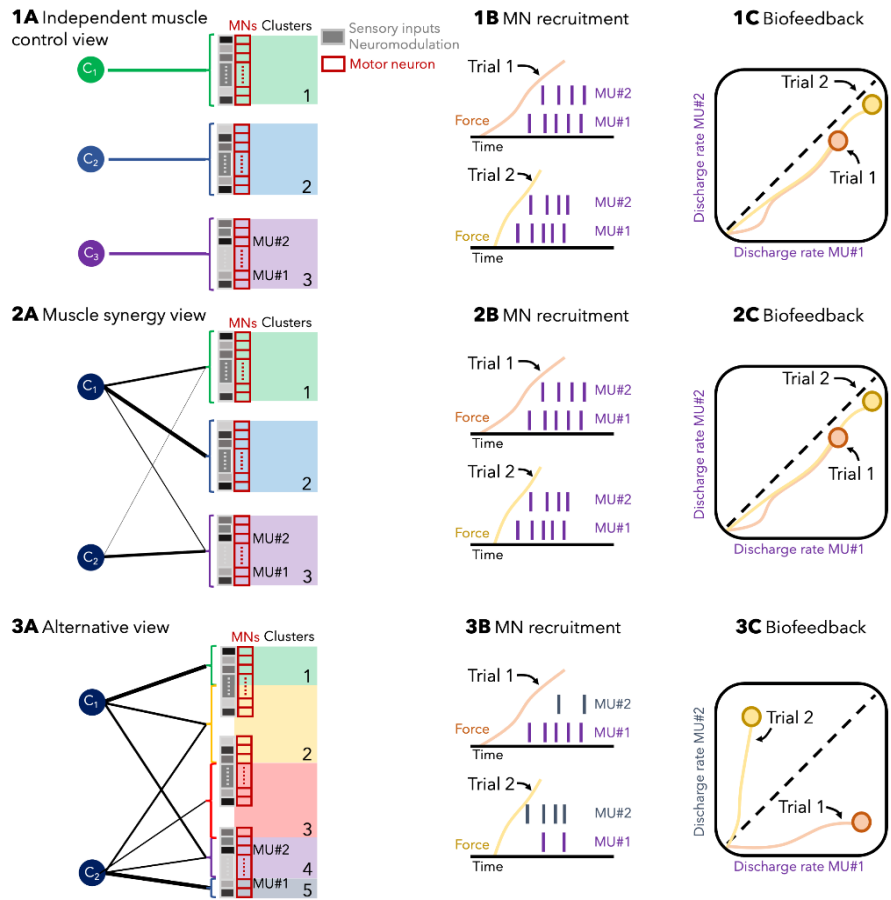

*Fig. 1. Distribution of common inputs to motor neurons. In the independent muscle control view (Panel 1A), the entire pool of motor neurons (MNs) receives a single common input ($C_i$). Within each pool, the recruitment follows the size principle, regardless of the motor task (Panel 1B). Due to this rigid control, a volitional independent control of motor neurons from the same pool cannot be achieved (Panel 1B and C). In the muscle synergy view (Panel 2A), the entire pool of motor neurons receives a set of common inputs with fixed weights, each common input being called a synergy. As for the independent muscle control view, the recruitment order follows the size principle within each pool, and activities of two motor neurons from the same pool cannot be independently modulated (Panel 2B and C). The framework that we propose (Panel 3) considers that common inputs are not projected to the entire pool of motor neurons. Instead, they are projected to clusters of motor neurons, which in turn project to the same or different muscles. As a consequence, we can observe reversal recruitment order between two motor neurons from the same pool if they receive different common input (Panel 3B). Moreover, volitional independent control of these motor neurons is theoretically possible (Panel 3C). It is noteworthy that motor neurons may receive specific proprioceptive feedback signals and that their intrinsic properties may independently change due to neuromodulation. Thus, pseudo flexible recruitment may be occasionally observed despite the presence of common inputs.*

This proposed view merges the concept of common input to motor neurons with the synergistic coordination of movement, together with the constraints imposed on the recruitment order. It is a relatively small change of view with respect to, for example, the muscle synergy theory, but we believe that this change is important in explaining many divergent observations, as we will now discuss. Importantly, this framework can be tested experimentally. For example, two important predictions of the framework are 1) motor neurons from the same pool (i.e. innervating the same muscle) may receive different common inputs and 2) motor neurons from different pools (i.e. innervating different, including distant muscles) may receive common inputs.

*3.1. Experimental evidence to support the control of functional groups of motor neurons*

Substantial synchrony or common drive has been shown to exist for most pairs of motor neurons innervating a muscle (Bremner *et al.*, 1991; Gibbs *et al.*, 1995), leading to the long-held belief that all motor neurons from a pool receive the same inputs (De Luca *et al.*, 1982; De Luca & Erim, 1994). However, few studies have considered that these common inputs may not be distributed over the entire pool of motor neurons that innervate a muscle. These latter studies mainly focused on muscles in which individual motor units are spatially organised within discrete neuromuscular compartments (English *et al.*, 1993). They observed that the recruitment of motor units from different regions may vary according to the mechanical constraints of the task (ter Haar Romeny *et al.*, 1984). Thus, evidence exists, mainly on multi-tendoned muscles, that synchronisation between motor neurons is stronger within than between different muscle compartments (Keen & Fuglevand, 2004; Reilly *et al.*, 2004; McIsaac & Fuglevand, 2007). Furthermore, results from a study by Madarshahian et al. (2021) support the notion that motor neurons from the flexor digitorum superficialis muscle form two groups that are controlled by two different (common) inputs. Recent studies have extended this observation to muscles that are not necessarily composed of neuromuscular compartments. The results of these studies suggest that despite most pairs of motor neurons from the same muscle receive common input, a significant proportion of them exhibit uncorrelated activity (Hug *et al.*, 2021; Tanzarella *et al.*, 2021; Del Vecchio *et al.*, 2022). For example, Del Vecchio et al. (2022) identified two independent neural synergies that controlled motor neurons innervating the vastus lateralis and vastus medialis muscles, but not all motor neurons innervating a muscle were controlled by the neural synergy mainly associated with that muscle. In other words, some motor neurons showed activity that correlated better with that of motor neurons in the other muscle than with the motor neurons in their "home" muscle. Similarly, Tanzarella et al. (2021) factorised the low-frequency oscillations of the discharge rate of motor neurons from 14 hand muscles to identify neural synergies. Although most of the motor neurons innervating the same muscle belonged to the same synergy, they again observed subgroups of motor neurons from the same muscle that belonged to different synergies. These observations are compatible with recent results obtained from both intramuscular and intracortical recordings made in macaque monkeys, where the motor neuron behaviour of the triceps brachii during a variety of motor tasks was best explained by multiple drives rather than a unique common drive to the motor neuron pool (Marshall *et al.*, 2022). Together, these results support the conceptual framework that motor neurons from the same pool do not necessarily receive the same synaptic input. Importantly, this evidence does not imply that some motor neurons from a certain pool are controlled through fully independent inputs with respect to all other neurons. Instead, it supports the claim that common inputs are projected onto functional groups of motor neurons rather than onto the entire pool.

Projections of inputs to functional groups of motor neurons may include projections to motor neurons from different pools. This is supported by several studies reporting common inputs distributed across muscles, including anatomically defined synergist (Gibbs *et al.*, 1995; De Luca & Erim, 2002; Mellor & Hodges, 2005) and non-synergist hand muscles. e.g. between extensor digitorum communis and flexor digitorum superficialis (Bremner *et al.*, 1991), between the flexor pollicis longus and flexor digitorum profundus (Hockensmith *et al.*, 2005). Furthermore, Hug et al. (2021) decoded the spiking activities of dozens of spinal motor neurons innervating six lower-limb muscles during an isometric multi-joint task. They identified subgroups of motor neurons that were partly decoupled from their innervated muscles. Specifically, subgroups of motor neurons from the same pool received different inputs, but shared common inputs with other subgroups of motor neurons innervating different, including distant, muscles. These results echo previous observations made by Gibbs et al. (1995) where the majority of participants exhibited synchronisation between the activity of motor neurons innervating the gastrocnemius and hamstring muscles. Although these studies could not determine the origin of the common synaptic input, their observations are compatible with the role of premotor interneurons (Levine *et al.*, 2014; Ronzano *et al.*, 2021) or corticospinal axons (Fetz & Cheney, 1980; Shinoda *et al.*, 1981) in the projection of correlated inputs to multiple motor neuron pools.

Together, these results imply that the CNS might control functional clusters of motor neurons rather than muscles. The proposed clustering of motor neurons based on common inputs has the main advantage of reducing the dimensionality of the control, while allowing flexibility in recruitment.

## 4. Implications for movement control

*4.1 Modular control*

The concept of synergistic control of movement has received considerable attention since its inception by Bernstein (1947). It is based on the assumption that grouping elements into stable groups reduces the number of variables to control and ultimately simplifies the production of movement. To date, the smallest unit of analysis within the synergistic model is the muscle, leading to the concept of *muscle synergy* (d'Avella & Bizzi, 2005; Ting *et al.*, 2015; Cheung & Seki, 2021). Muscle synergies are identified by factorising interferential electromyography (EMG) signals from multiple muscles. This approach inherently constrains the dimensionality of the neural control to be less than or equal to the number of recorded muscles and relies on the underlying assumption that all motor neurons from a pool receive the same inputs, that is, the same common inputs with the same weight for the entire pool (Fig. 1B). Indeed, the activation signal of the synergies is often represented as a projection to entire motor neuron pools (e.g., Fig. 2 in Cheung et al., 2021; Fig. 1 in Giszter, 2015). However, as proposed in our framework and supported by experimental data (Section 3), common inputs would not be projected to motor nuclei innervating muscles, but across nuclei, partly irrespective of muscle innervation.

Grouping motor neurons into functional clusters might provide functional advantages. First, it reduces the initial large dimensionality of spinal motor neurons by grouping them into a smaller number of clusters. Control dimensionality is further reduced by the distribution of common inputs across clusters, which implies that the number of control signals for a given task may be smaller than the number of clusters. Notably, a reduction in dimensionality with respect to the number of motor neurons can also be achieved with common inputs distributed to motor neuron pools, as proposed by the classical views (Figs. 1A and B). However, the combination of functional clusters, as proposed by our framework, allows for a more flexible system relative to the classic muscle synergy control, where the pools of motor neurons rigidly receive the same inputs. By grouping motor neurons into functional clusters, the CNS can independently control motor neurons from the same muscle to comply flexibly with concurrent task constraints. For example, Hug et al. (2021) analysed an isometric multi-joint task that required the combined action of the gastrocnemii and hamstring muscles to extend the lower limb (Cleather *et al.*, 2015) together with an action of the gastrocnemii muscles to orientate the output force at the foot. The observed organisation of motor neuron clusters reflected task demands; for example, it included both a cluster of motor neurons innervating the gastrocnemius and hamstring muscles and a cluster innervating the lateral and medial gastrocnemius muscles (Hug *et al.*, 2021). Such an organisation may allow the CNS to independently control two motor actions, while reducing control dimensionality. In other words, these functional clusters may be recruited as functional units to control isolated knee flexion, isolated plantar flexion, or combined knee flexion and plantar flexion. This functional advantage is not restricted to the control of the biarticular muscles. For example, the lateral and medial heads of the quadriceps share two main functions: producing knee extension torque and controlling the patellofemoral joint (Lieb & Perry, 1968). The fact that the motor neurons from each of these muscles are not necessarily controlled by the same common input (Del Vecchio *et al.*, 2022) may allow flexibility in independently controlling these important actions.

Further work is required to explore the robustness of this motor neuron grouping across different behaviours or mechanical constraints. Notably, this view of motor control modularity implies that conventional muscle synergy analysis is not appropriate for identifying the control dimensionality, while an analysis at the motor neuron level is necessary.

*4.2 Recruitment order*

Since the initial formulation of Henneman's size principle, notable exceptions have been reported, with a reversed order of recruitment (Desmedt & Godaux, 1981; Kishore *et al.*, 2014; Azevedo *et al.*, 2020; Marshall *et al.*, 2022) or volitional independent control of motor units (Basmajian, 1963; Formento *et al.*, 2021). Although these divergent observations are incompatible with the classical view of the size principle applied to the entire pool of motor neurons innervating a muscle, we contend that they can be explained by changing the scale at which the size principle is considered.

The relationship between the excitation threshold of a motor neuron and its size depends on underlying biophysical constraints (Caillet *et al.*, 2021). Therefore, challenging the Henneman's size principle is equivalent to challenging the presence of common input. The framework presented in Section 3 implies that the size principle applies to clusters of neurons that receive common inputs, rather than to motor neuron pools. Therefore, a reverse recruitment order can be observed between motor neurons from the same pool but from different clusters. It significantly differs from the interpretation that the activity of these motor neurons can be selectively modulated by the CNS (Basmajian, 1963; Formento *et al.*, 2021). Specifically, a change in the order of recruitment between motor neurons belonging to different clusters may be achieved by changing the relative strength of the common inputs to the clusters. This was previously hinted at with perfect clarity by Bawa et al. (2014): "*[…] when discussing recruitment order, the motoneuron pool should be operationally defined as the group of motoneurons that receive excitatory synaptic input to drive the functional movement, not the pool of motoneurons defined by anatomy. The validity of the size principle should then be evaluated within this operationally defined motoneuron pool to determine if recruitment proceeds from small to large*". Notably, further flexibility in recruitment within a functional cluster of motor neurons may be achieved through specific proprioceptive feedback signals and selective modulation of the intrinsic properties of motor neurons.

## 5. Conclusion and future directions

We propose a conceptual framework of the neural control of movement, which merges the concept of common input to motor neurons and synergistic coordination of movement, together with the constraints imposed by recruitment order. A central feature of this framework is the distribution of common inputs to clusters of motor neurons, which partly overlap with the muscle innervation. Such a framework marks a transition from muscle synergy theory to motor neuron synergy theory. It is, however, important to note that the structure of common inputs proposed in Fig. 1C may be more complex, such that clusters may partly overlap, allowing further flexibility in recruitment strategies, which leads to the important open question of the level of flexibility in motor neuron clusters. This question should be addressed through experiments performed on a vast repertoire of natural behaviours or virtual tasks decoupled from mechanical constraints. The results of such experiments may either confirm the proposed framework and may add new features to enrich it, or on the contrary, may disprove some parts of this view.